\documentclass[prl,amsmath,amssymb,twocolumn,floatfix]{revtex4-2}
\usepackage{graphicx}
\usepackage{dcolumn}
\usepackage{bm}
\usepackage{color}

\begin{document}

\preprint{APS/123-QED}

\title{Reconfigurable Elastic Metamaterials: Engineering Dispersion with Meccano\textsuperscript{\texttrademark}}

\author{G.~J.~Chaplain}
\email{g.j.chaplain@exeter.ac.uk}
\author{I.~R.~Hooper}
\author{A.~P.~Hibbins}
\author{T.~A.~Starkey}
\affiliation{Centre for Metamaterial Research and Innovation, Department of Physics and Astronomy, University of Exeter, Exeter EX4 4QL, United Kingdom}


\begin{abstract}
 We design, simulate and experimentally characterise a reconfigurable elastic metamaterial with beyond-nearest-neighbour (BNN) coupling. The structure is composed from the popular British model construction system, Meccano\textsuperscript{\texttrademark}. The Meccano\textsuperscript{\texttrademark} metamaterial supports backwards waves with opposite directions of phase and group velocities. We experimentally verify three distinct configurations and acoustically infer their spatial vibration spectra.
\end{abstract}

\maketitle

Dispersive waves whose velocity varies non-linearly with frequency are of great interest to the wave-physics community. Metamaterials are structured materials with often periodic structuring that are capable of supporting such waves that owe their dispersive properties due to their geometric design. Metamaterial concepts were originally explored in the electromagnetic domain \cite{veselago1967electrodynamics,pendry1999magnetism,pendry2000negative,smith2000composite}, but quickly saw the ideas translated to other wave systems, from acoustics to elasticity \cite{craster2012acoustic,zhou2012elastic,hussein2014dynamics,kadic20193d}, and to other periodic systems such as photonic and phononic crystals \cite{luo2002all,foteinopoulou2005electromagnetic}. Engineering dispersive wave properties is of particular importance in mechanical systems where vibration isolation and energy harvesting devices rest on manipulating elastic wave propagation; elastic metamaterials offer the promise to unlock such applications. Such devices span orders of magnitude in length scale, from microscale surface acoustic wave devices \cite{kahler2022surface} and mesoscale bench-top devices \cite{colombi2017enhanced,chaplain2020topological}, all the way up to macroscale seismic metamaterials \cite{colombi2016seismic,pouya2021sub} for groundborne vibration control. Backwards waves are one class of dispersive waves that possess unique properties; their phase and group velocities are antiparallel, and as such the propagation of wave-fronts is in the opposite direction to the energy flow - a phenomenon well-known to electrical engineers since the early 1960s \cite{clarricoats1960non}, and that naturally occurs in certain settings in elasticity, such as for Lamb modes in anisotropic plates and axisymmetric modes in pipes \cite{mindlin1957mathematical,shuvalov2008backward,tamm2017negative,chapman2021wronskian}. Recent advances in elastic and acoustic metamaterials \cite{chen2021roton} have achieved analogous backwards wave effects that can be bespoke and dictated by their structure, offering a route to tailoring dispersion by introducing reconfigurable components. 

Metamaterial design and functional interpretation is often predicated on the analysis of wave propagation in periodic structures \cite{maznev2015waveguiding,matlack2018designing}. The ubiquity of such systems has inspired similar analytical techniques to be adopted to describe wave phenomena varying from the crystal scattering of X-rays, thermal vibrations in crystal lattices, electronic motion in metals, and analogous electric engineering systems such as electromagnetic wave propagation in periodic circuits \cite{brillouin53a, eleftheriades2002planar}. Indeed this analysis can be traced back to the ``delightfully written'' \cite{born1946wave} seminal work of L\'{e}on Brillouin \cite{brillouin53a}, who devotes a mathematical treatment to wave propagation in periodic structures, founded on a simple coupled resonator model in the form of particles coupled by elastic springs. The introduction of Brillouin Zones (BZ), and their relation to the reciprocal lattice, forms the foundation for interpreting the dispersion, i.e. the frequency-wavevector relation $\omega(\boldsymbol{k})$, of waves supported within periodic structures; the overarching theme connecting the disparate areas of physics described by this theory is that the associated periodic boundary conditions are satisfied by Bloch-Floquet waves that provide elegant solutions in the framework that the reciprocal lattice and associated Irreducible Brillouin Zone (IBZ) provide.

The canonical introductory model for periodic structures is a mechanical, one-dimensional, infinite chain of identical masses coupled to nearest-neighbours (NN) by identical springs. This is now commonplace in undergraduate courses, and indeed is the starting model of Brillouin \cite{brillouin53a}. This exemplar system exhibits a non-linear dispersion relation, a cut-off frequency for systems of equal particles, and pass and stop-bands for systems of different particles. The dispersion relation can be altered by including BNN coupling; as discussed by Brillouin, for interactions extending to the $L$th nearest neighbour, the form of the dispersion relation \cite{brillouin53a}
\begin{quote}
    ``[...] will be expressed as a polynomial of degree $L$''.
\end{quote} 
And so, as described by Brillouin, for interactions with the $L$th nearest neighbour, there will be $L-1$ extrema within the first Brillouin Zone, resulting in regions with altering signs of the group velocity, $\boldsymbol{v}_g \equiv \frac{\partial\omega}{\partial \boldsymbol{k}}$. For the case of NN-only interactions the singular extrema occurs at the band edge owing to the formation of standing waves via the Bragg condition. 

\begin{figure}[t]
    \centering
    \includegraphics[width = 0.45\textwidth]{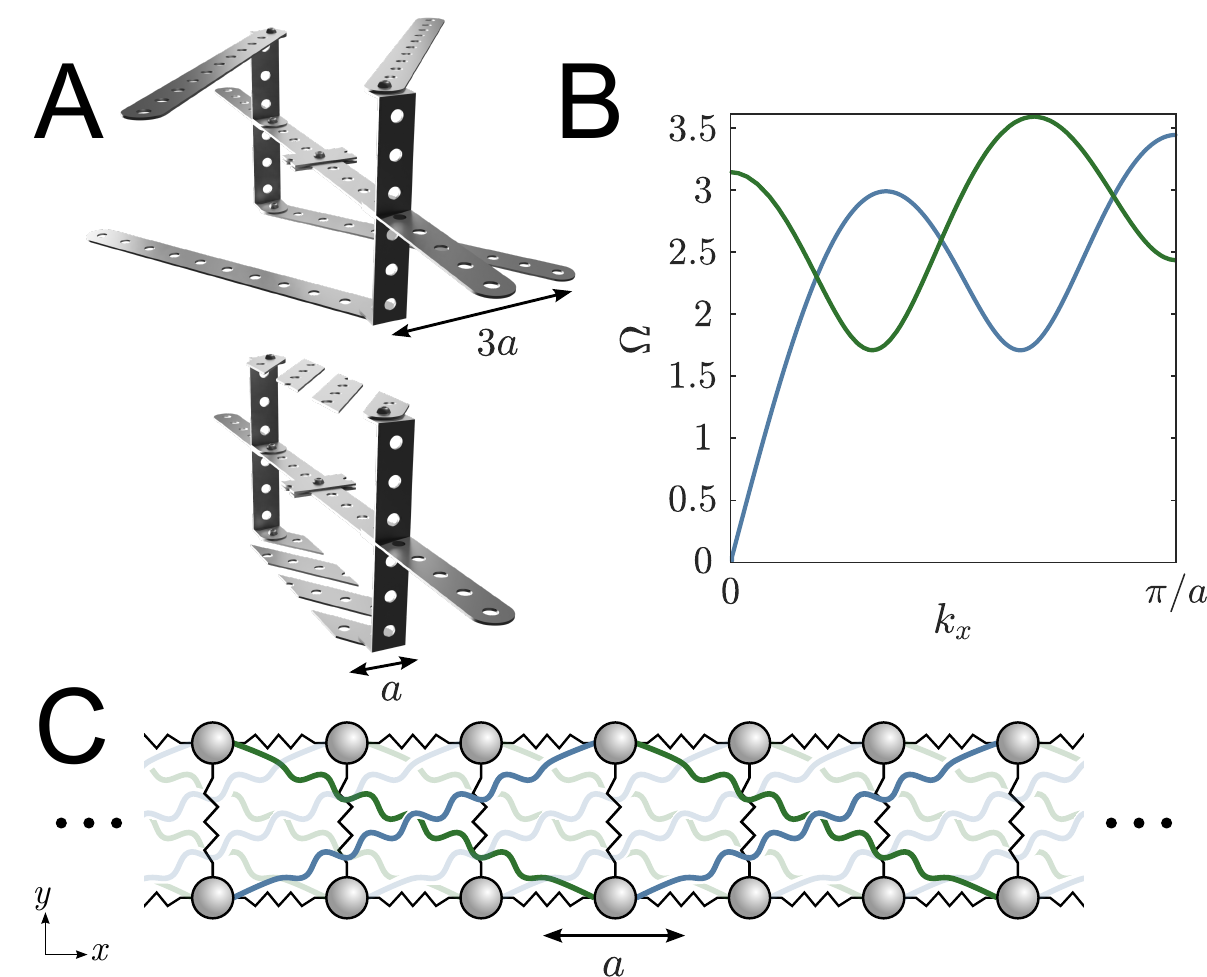}
    \caption{(A) Meccano\textsuperscript{\texttrademark} geometry: upper figure shows the unit cell with BNN connections, with the lower figure showing the periodically repeating unit cell of width $a$. (B) Example dispersion curves with multiple extrema within first BZ, obtained from a coupled resonator toy model (C). Toy mass-spring model with BNN coupling (green and blue springs).}
    \label{fig:Toy}
\end{figure}

BNN coupling is not unique to metamaterials or other periodic analogues. Indeed it is commonplace to account for additional (sometimes `non-local') interactions when considering substantially more complex electronic band structures, where extra couplings arise in, for example, tight-binding models \cite{hill2009dielectric,neto2009electronic}, $\boldsymbol{k}\cdotp \boldsymbol{p}$ perturbation theory \cite{luttinger1955motion} and density functional theory \cite{hohenberg1964inhomogeneous,kohn1965self}. 

A coalescence of the ideas surrounding BNN interactions has recently been realised for acoustic and elastic metamaterials and, specifically, next-next-nearest-neighbours i.e. $L = 3$ \cite{chen2021roton,iglesias2021experimental} that have also been described as non-local metamaterials \cite{wang2022nonlocal}. The intricately designed structures of \cite{chen2021roton} interconnect one unit cell with a physical join to one three unit cells away, providing an additional degree of interaction and an alternative channel for power-flow. The maxima and minima of the modal dispersion within the first Brillouin Zone are associated with these competing channels, resulting in a characteristic dispersion relation that shows some analogies to that of rotons \cite{chen2021roton}. 

Competing power channels, leading to regions of negative dispersion, can be achieved by other means than physically joined BNN coupling in acoustics and elasticity. Indeed in electromagnetic metamaterials, such as the Sievenpiper `mushroom array' \cite{sievenpiper1999high}, there are no geometrical elements that couple beyond the fundamental cell, but inherent competing power channels above (air) and below (hyperbolic media, vias)  the structured surface result in regions of negative dispersion within the first Brillouin Zone. Extensions of this are well documented in electromagnetism for both bulk and surface waves \cite{shin2006all,lezec2007negative,yao2008optical, dockrey2016direct,tremain2018isotropic}, and for waveguide modes \cite{seetharaman2018realizing}. Similar tailored dispersion relations can arise from other mechanisms, such as chiral-induction in micropolar elastic materials \cite{kishine2020chirality}, symmetry breaking \cite{chaplain2020delineating,Ward2022} and topology \cite{chen2014nonlinear,xin2020topological}.

In this paper, taking inspiration from the recent `roton-like' metamaterials, we design, simulate and experimentally characterise a modular, reconfigurable elastic metamaterial with beyond nearest neighbour coupling. We overcome the "serious experimental challenge" \cite{fleury2021non} surrounding the decoupling of NN and BNN connections by designing an elastic metamaterial that is easily reconfigurable using the popular British model construction kit Meccano\textsuperscript{\texttrademark}, owing to its modular, cost effective design (and partly due to childhood nostalgia). For over a century Meccano\textsuperscript{\texttrademark} has been the first opportunity for aspiring young scientists and engineers to explore aspects of mechanics by designing, building and playing with Meccano\textsuperscript{\texttrademark} sets, which typically contain building blocks such as strips, plates, and angle girders, with additional components such as wheels, axles and gears. We opt to design in earlier-style steel Meccano\textsuperscript{\texttrademark}, rather than the modern day plastic equivalent, due to its elastic material properties. As such the dimensions are quoted in imperial units (inches ") as in the original designs \cite{meccano}.

\section{Meccano\textsuperscript{\texttrademark} Metamaterials: Design}

\begin{figure*}[t]
\centering
\includegraphics[width=0.9\linewidth]{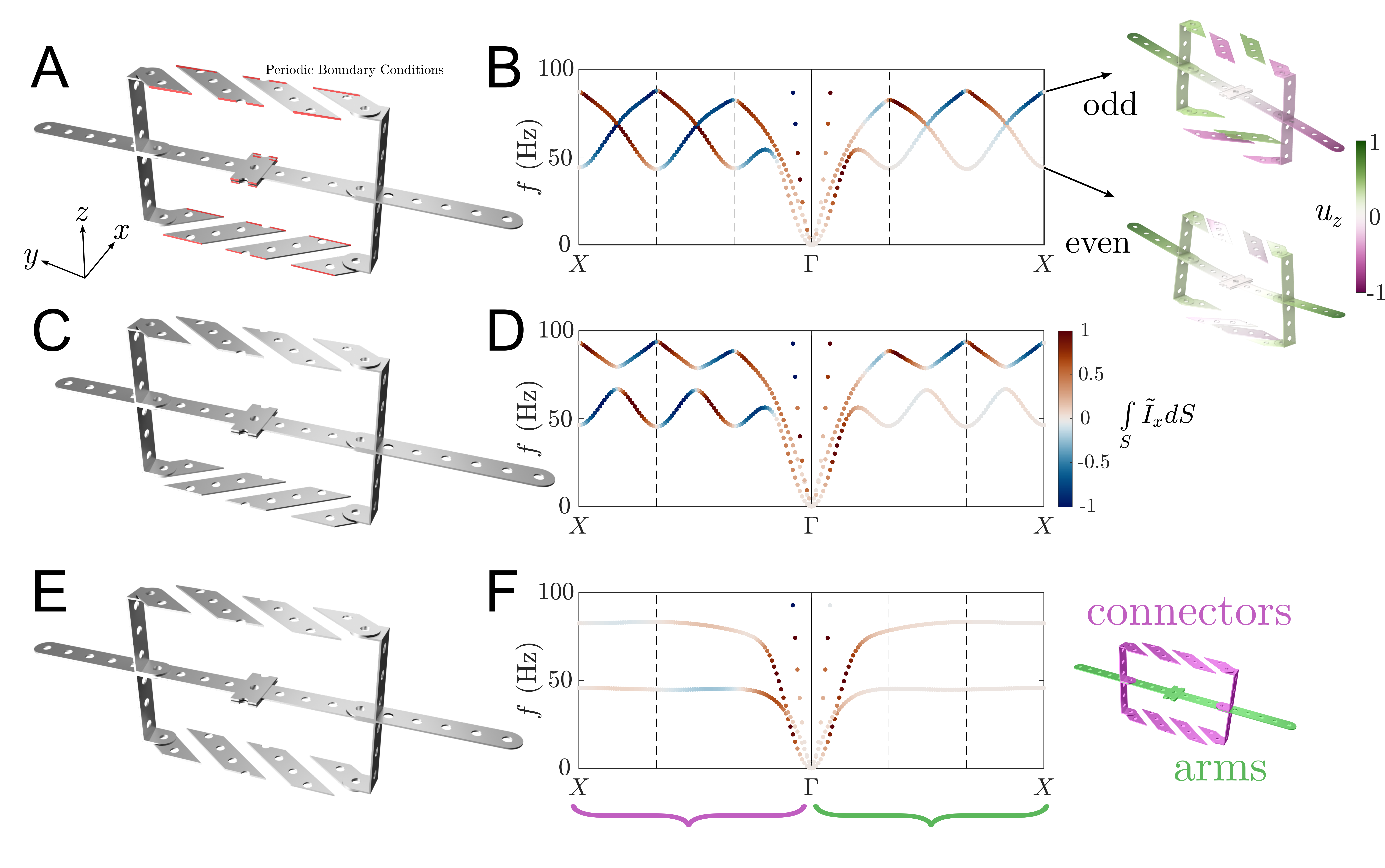}
\caption{The three unit cell configurations and their respective dispersion curves: (A) Schematic of the unit cell for the symmetric configuration, with edges highlighted in red where periodic boundary conditions have been employed in FEM modelling \cite{comsolSolidMech}. (B) Dispersion curves obtained via frequency domain eigenvalue simulations. The colourscale depicts the value of the normalised integrated mechanical energy flux in the direction of the periodicity, $\int_{S}\tilde{I}_x dS$, over the regions $S$ which denote either the connectors (left hand side of dispersion plot) and arms (right hand side of dispersion plot). At the band edge the eigenmode-shapes are shown, with colourscale showing the normalised out-of-plane displacement, $u_z$; the displacement of the perpendicular arm is purely odd or even in $z$. The dashed lines indicate $k_x = \pm\pi/3a$ and $k_x = \pm2\pi/3a$. (C,E) Schematics of the unit cells for the asymmetric and parallel configurations respectively. (D,F) Dispersion curves, as in (B), but for the asymmetric and parallel configurations respectively. Shown next to (F) is a schematic of the regions $S$ (the connectors and arms) over which the complex mechanical energy flux has been integrated.}
\label{fig:disp}
\end{figure*}

\begin{figure*}[t]
\centering
\includegraphics[width=.75\linewidth]{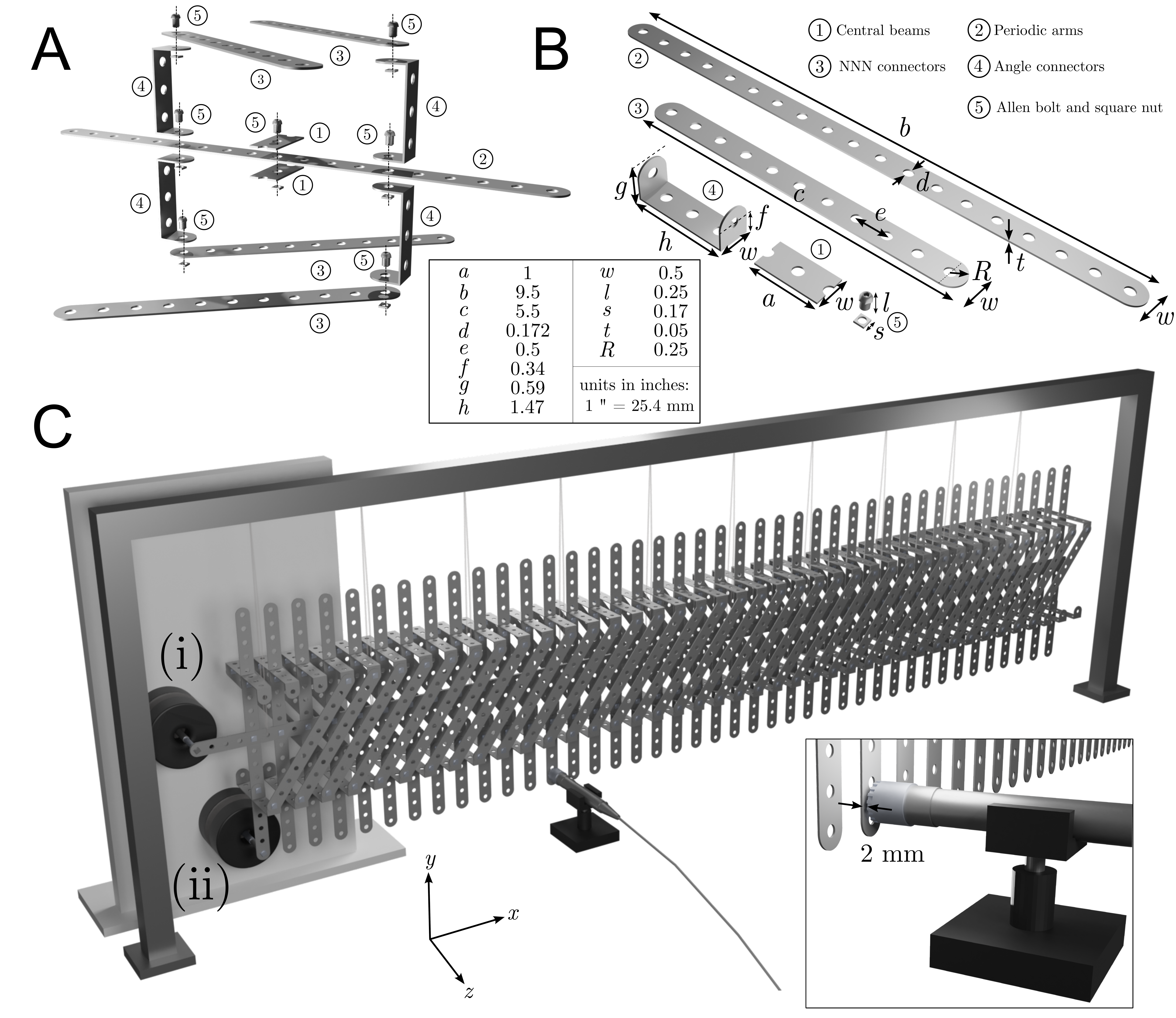}
\caption{Schematic diagrams. (A) instruction-manual style diagram for constructing the unit cell (symmetric configuration example). (B) Meccano\textsuperscript{\texttrademark} pieces required for the unit cell. Note the central beam shown is periodically repeated and forms a whole beam the entire length of the structure. (C) Experimental setup for acoustically characterising the dispersion curves of the structure. Full details in Materials and Methods section.}
\label{fig:Setup}
\end{figure*}

To facilitate the design of an elastic metamaterial with BNN coupling, we turn to a canonical mass-spring toy model, akin to Brillouin \cite{brillouin53a}. The toy model is composed of two coupled chains of dipole-like masses, linked to nearest neighbours by springs, with additional BNN coupling to the third nearest neighbour, a schematic of which is shown in Fig.~\ref{fig:Toy}(C). The two supported modes have $L - 1 = 2$ extrema within the first BZ, forming a pair of orthogonal modes as shown in Fig.~\ref{fig:Toy}(B). Figure~S1 details the masses and springs constants used to formulate the governing difference equation and ensuing eigenvalue problem as detailed in the supplementary information. The toy mass-spring model allows intuition to be gained on the influence of symmetry breaking on the dispersion relations of the modes supported by the chains, which we investigate through our proposed reconfigurable Meccano\textsuperscript{\texttrademark} structure. In Fig.~\ref{fig:Toy}(A) we highlight the unit cell of one of the possible configurations of the Meccano\textsuperscript{\texttrademark} metamaterial, resulting from extension of the toy model.

Our Meccano\textsuperscript{\texttrademark} designs are built from cold rolled steel (see Materials and Methods), and comprise two key components to introduce coupling on different length-scales within the structure; (1) a central strip that runs the length of the sample, to which are attached perpendicular beams, and (2) inter-unit-cell connecting strips that form a scaffold above and below the central strip, making use of a convenient Pythagorean triple that fits the discrete holes available on the Meccano\textsuperscript{\texttrademark} designs, each spaced by 0.5". The central beam provides nearest-neighbour coupling between unit-cells (of width $a$), of which the perpendicular arms are the main resonant element, and the connectors provide BNN coupling. This design bares an equivalence to the toy model: the perpendicular arms and angled joints to the connectors approximate lumped masses connected by the vertical springs in the toy model. The central beam forms the spring-like connection to adjacent cells in each chain of masses, with the connectors providing the analogy to the long-range springs as shown in Fig.~\ref{fig:Toy}(C). The lowest two bending modes of the perpendicular arms, where the displacement is predominantly in the $z$ direction, then possess the same symmetries as the two eigenmodes present in the toy model. 

The structure can be configured in a number of ways to tailor the mode dispersion. We demonstrate three configurations, that are easily built and re-built, and support modes with distinct dispersive characteristics. We obtain the dispersion curves for each configuration using the Finite Element Method (FEM) \cite{comsolSolidMech}, solving the 3D equations of elastodynamics in the frequency domain. We define two of the three configurations in relation to the symmetry of the perpendicular arm with respect to the $z-x$ plane through the central beam and term them (i) the symmetric configuration (Fig.~\ref{fig:disp}(A)), where the perpendicular arm extends symmetrically in $y$ from the central beam, and the connectors above and below the central beam are crossed; (ii) the asymmetric configuration (Fig.~\ref{fig:disp}(C)), where the perpendicular arm is displaced so that the arm is longer on one side of the connectors (and therefore asymmetric about the central beam). Configuration (iii) returns to a symmetric arrangement of the perpendicular arm, but now with the connectors above and below the central beam being parallel. A schematic of all parts is shown in Fig.~\ref{fig:Setup}(A-B), detailing the conventional Imperial units of classic British Meccano\textsuperscript{\texttrademark}. The central beam section (component 1) is periodically repeated along the axis of the sample, in the $x$-direction. 

The lowest eigenfrequencies of each configuration are evaluated by employing Floquet-Bloch periodic boundary conditions in the FEM model, and solving the ensuing eigenvalue problem. The periodic boundaries are highlighted as red-edges in Fig.~\ref{fig:disp}(A). Figures~\ref{fig:disp}(B,D,F) show the corresponding dispersion curves for the three configurations (i,ii,iii) respectively. The colour-scale in each dispersion diagram represents the normalised integrated complex mechanical energy flux (in the direction of periodicity) over two regions: the left half of the dispersion curves shows the integrated complex mechanical energy flux over the connectors, as highlighted in magenta in Fig.~\ref{fig:disp}(F), while the right half shows the same calculation over the perpendicular arms, as highlighted in green in Fig.~\ref{fig:disp}(F). This quantity provides a relative measure of the power flow along the different channels provided by the short- and long-range couplings that co- and counter-propagates, demonstrating that is these competing power flows that drives the extrema in the dispersion curves. Of course, the net power flow must always be away from a given source; in Fig.~S3 we show the normalised sum of the complex mechanical energy fluxes over the two channels (arms and connectors), confirming the net power is always in one direction.

Each of the three configurations possess unique dispersive properties that are related to the internal symmetries possessed by the unit cell. As we consider the lowest two bending eigenmodes, we describe the modal symmetries in terms of the out-of-plane displacement ($u_z$) of the perpendicular arms. In the symmetric configuration, the structure supports an orthogonal pair of eigenmodes \cite{makwana2018geometrically} that manifest in the form of degeneracies in the band structure, i.e. at the points where the dispersion curves of the two modes cross. The out-of-plane displacement of the perpendicular arm is (anti)symmetric, giving rise to the notion of (odd)even modes, as shown by the out-of-plane displacement fields in Fig.~\ref{fig:disp}(B) at the band edge where the wavevector $k_x = X \equiv \pi/a$; standing waves form here and the energy flux along the structure vanishes. The orthogonality of these modes suggests they can be excited independently or mixed by utilising a forcing that matches the modal symmetry (see Results).

The asymmetric configuration breaks the reflectional symmetry of the perpendicular arm about the $z-x$ plane defined through the central beam, thereby lifting the degeneracies present in the dispersion curves for the symmetric configuration. This is shown in Fig.~\ref{fig:disp}(D), where the mixing of the modes can be seen when comparing the energy fluxes of the symmetric and antisymmetric modes along the arms, resulting in anti-crossings between the positively and negatively dispersing modes. We show the eigenmode shapes where the out-of-plane displacement of the perpendicular arm is neither odd nor even in Fig.~S3.

In the parallel configuration, the perpendicular arms are once again symmetric about the about the $z-x$ plane defined through the central beam, but rotational symmetry about the $y$-axis is broken. The resulting dispersion curves are much flatter, although still retaining the same number of extrema within the IBZ. This orientation of connectors biases the direction of power flow along them, as seen by the integrated mechanical energy fluxes of the two channels that remains predominantly aligned. 

Having demonstrated the reconfigurable nature of the unit cells, and the effects they have on the dispersion of the supported waves on an infinitely repeating system, we now present the experimental characterisation of each configuration and compare with FEM simulations of the finite structure.

\section{Results}

\begin{figure*}
\centering
\includegraphics[width=.85\linewidth]{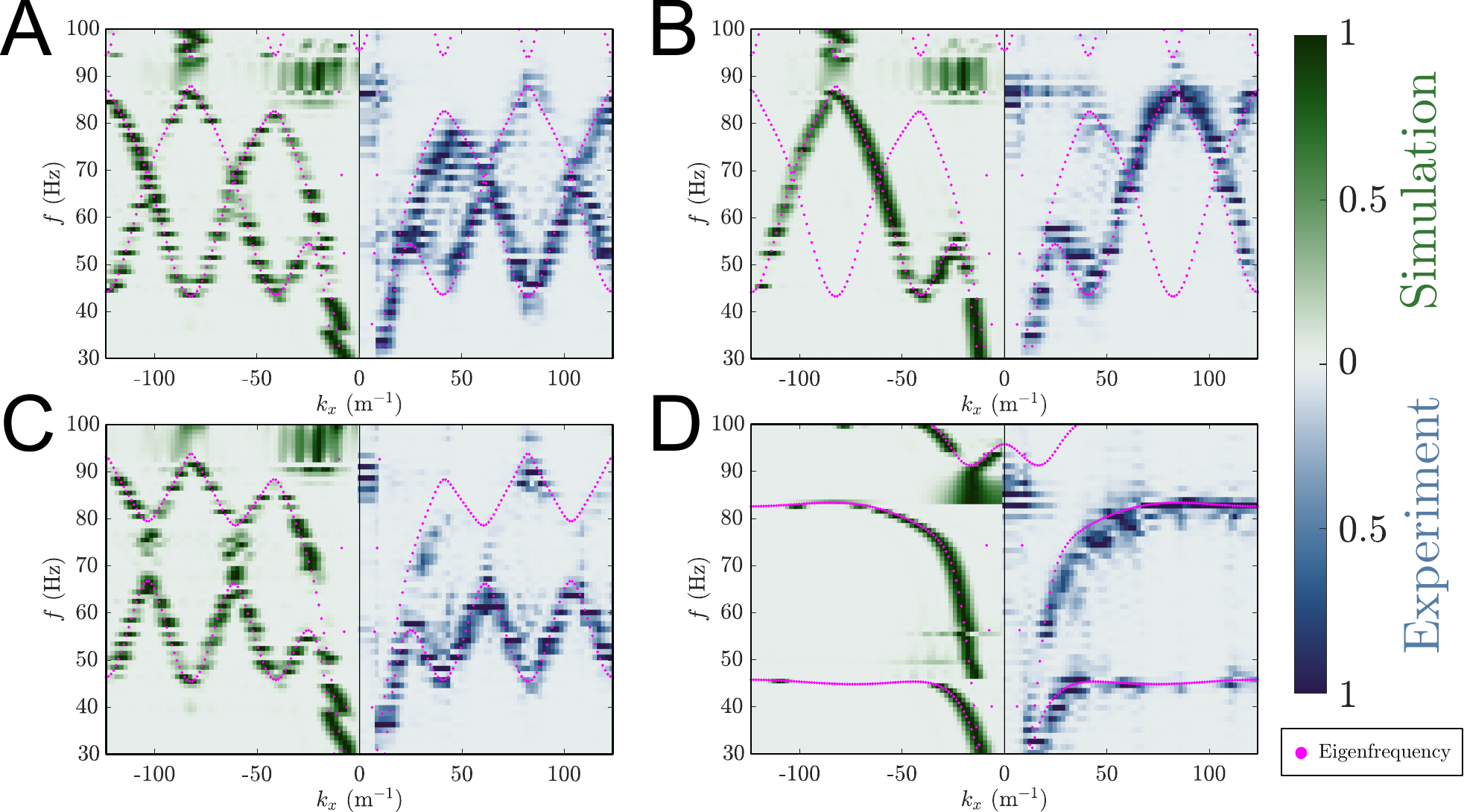}
\caption{Comparisons between numerical (FEM) and experimental results. (A) Normalised, per frequency, Fourier amplitude of numerically (left hand side) and experimentally (right hand side) determined dispersion spectra, for the symmetric configuration excited in position (ii) from 30-100 Hz in 1 Hz steps. Overlaid are the eigenfrequency solutions from the infinitely periodic model (Fig.~\ref{fig:disp}). (B) As for (A), but for excitation in position (i), therefore isolating the even mode. (C) As for (A), but for the asymmetric configuration. (D) As for (A), but for the parallel configuration. }
\label{fig:Results}
\end{figure*}

Owing to the scarcity of original Meccano\textsuperscript{\texttrademark} pieces, and the variations in its composition and material properties, a custom made Meccano\textsuperscript{\texttrademark}--like set was fabricated following the original Meccano\textsuperscript{\texttrademark} dimensions \cite{meccano}. The set was laser cut from a single sheet of cr4 (cold rolled s275 equivalent), and assembled into the unit cells following the schematic in Fig.~\ref{fig:Setup}(A). Figure~\ref{fig:Setup}(C) shows the full structure, in the symmetric configuration, consisting of 41 unit cells, with an extended central beam to which a shaker (PASCO Mechanical Wave Driver 317
model number 2185) is attached at either position (i) or (ii), depending on the desired excitation condition. The structure is suspended from a rigid frame using nylon wire at equally spaced points long the beam length (as shown in Fig.~\ref{fig:Setup}(C)). We adopt a simple  experimental set-up to measure the vibrations acoustically. The structure was driven to a steady state at discrete frequencies and a microphone was manually moved and spaced $2$ mm from each perpendicular arm along the array, as shown in Fig.~\ref{fig:Setup}(C). The local pressure fluctuations in the air generated by the vibrating arms was measured by the microphone, and the dispersion spectra recovered by spatio-temporal Fourier analysis of the data by means of the Fast Fourier Transform (FFT). The experimental results are shown in Fig.~\ref{fig:Results} on the right side of each plot; the left side of each plot shows the results from frequency domain FEM simulations that mirror each of the experimental configurations: the same finite number of unit cells are simulated with corresponding forcing conditions at the shaker positions, and the out-of-plane displacement is extracted along the perpendicular arms and analysed by a spatial FFT to obtain the dispersion spectra along the Meccano\textsuperscript{\texttrademark} device, as shown in Fig.~S4.

Figure~\ref{fig:Results} shows the excellent agreement between the numerically simulated and experimental dispersion curves, normalised on a per-frequency basis. Figure~\ref{fig:Results}(A) shows comparisons of the numerical and experimental dispersion for the symmetric configuration excited in position (ii) as shown in Fig.~\ref{fig:Setup}(C), which contains a mixture of the even and odd modes shown in Fig~\ref{fig:disp}(B). Figure~\ref{fig:Results}(B) shows a similar comparison but for excitation at position (i); this only drives modes with an even out-of-plane displacement of the perpendicular arms as the source is aligned with the Meccano\textsuperscript{\texttrademark} axis. Figures~\ref{fig:Results}(C,D) show the comparisons for the asymmetric and parallel configurations respectively. In all four plots the eigenfrequency results for the infinitely periodic case are overlaid (Fig.~\ref{fig:disp}). For the case of the asymmetric configuration (Fig~\ref{fig:disp}(C)), the shaker and measurement position was on the longer side of the asymmetric perpendicular arm  (Fig.~S4). 

\section{Discussion}
The dispersion characteristics of elastic metamaterials and other periodic structures are conventionally fixed once the sample has been fabricated. Here we developed a novel, reconfigurable elastic metamaterial that leverages beyond-nearest neighbour interaction, capable of supporting backwards waves. Qualitative analogies to a canonical mass-spring model motivated the design paradigm surrounding the symmetries of the structure. The BNN coupling provides alternate channels for power flow, the competition between which results in extrema within the first Brillouin Zone, and as such unlocks motivation for vibration sensing applications in, for example, scaled down surface acoustic wave devices. 

The device was inspired by the popular model construction kits, Meccano\textsuperscript{\texttrademark}, and motivates the development of analogous modular metamaterials with replaceable components in addition to serving as a tactile, innovative pedagogical method for introducing beyond nearest neighbour interactions in metamaterials and phononic crystals. The acoustic experimental characterisation of the structure, by measuring the local pressure fluctuations induced by the resonant elements of the structure, provides an alternate method for vibration characterisation on the lab-bench scale using simple acoustic methods, where more complex and expensive methods, (e.g. laser velocimetry) may not be available.

\section{Materials and Methods} 

\noindent{\textbf{Numerical Simulations}:}
The dispersion curves (Fig.~\ref{fig:disp}) of each configuration of the Meccano\textsuperscript{\texttrademark} were obtained via the FEM model using an eigenfrequency study in the Solid Mechanics module within COMSOL Multiphysics (version 6.0). The dispersion spectra shown in Fig.~\ref{fig:Results} was obtained through a frequency domain simulation of 41 unit cells of the Meccano\textsuperscript{\texttrademark} metamaterial (in vacuo) with free boundary conditions on all boundaries except the position of the shaker, which was modelled as a boundary load at positions (i) and (ii), as in Fig.~\ref{fig:Setup}(C). The out-of-plane displacement was then extracted along a line down the length of the sample and analysed by means of the FFT. A schematic of the computational domain and boundary conditions is shown in Fig.~S4. The mild steel mechanical material properties of the Meccano\textsuperscript{\texttrademark} were assumed as; density $\rho = 7870$ kg/m$^3$, Poisson's ratio $\sigma = 0.287$, and elastic modulus $Y = 310$ GPa.

\noindent{\textbf{Meccano\textsuperscript{\texttrademark} components}:} A Meccano\textsuperscript{\texttrademark} set was designed following the original Meccano\textsuperscript{\texttrademark} dimensions set out in the `Liverpool' engineering drawings \cite{meccano}. The Meccano\textsuperscript{\texttrademark} components were laser cut by Luffman Engineering (Tiverton, UK), from a single, 1.2 mm thick, mild steel sheet (cold rolled (CR4)).

The sample was assembled, as shown in the manual-style schematic in Fig.~\ref{fig:Setup}(A), with original Meccano\textsuperscript{\texttrademark} bolts (37b Allen Bolt $1/4$ (6mm) Zinc) and corresponding square nuts.

\noindent{\textbf{Experimental set-up and method}:} A schematic of the experimental set-up is shown in Fig.~\ref{fig:Setup}(C). The Meccano\textsuperscript{\texttrademark} sample was hung horizontally from a bar using Nylon fishing wire looped through the central strip that runs the length of the sample. The sample was excited using a mechanical shaker (PASCO Mechanical Wave Driver model number 2185) driven by a variable frequency signal generator (Agilent 33220A Arbitrary Waveform Generator). Two excitation positions were considered (i) and (ii) (Fig.~\ref{fig:Setup}(C)) to provide either a symmetric displacement excitation (in (i)), or a displacement with mixed symmetry (ii) in order to demonstrate the orthogonal nature of the modes in the symmetric configuration. For all measurements the shaker was driven by a $\pm 5$ V peak to peak sine wave.

To measure the displacement of the vertical beams, the local pressure fluctuations in the air at each vibrating arm were recorded. At each arm, a microphone (Brüel \& Kjær Type 4940 $1/2$-in Free-Field Microphone, with Pre-conditioning amplifier) was positioned 2 mm from the sample so that the head was centered in the width of the Meccano\textsuperscript{\texttrademark} bar horizontally and between the two lower holes vertically (as shown in Fig.~\ref{fig:Setup}(C) inset) and the signal recorded on an oscilloscope (Siglent  SDS2352X-E). Acoustic data was recorded with a sampling frequency of 500 kSa s$^{-1}$, for a total time of 0.28 sec, and with 16 averages.

The experimental procedure was as follows; the microphone was positioned at the first arm, the structure was driven to a steady state at discrete frequencies (30 to 100 Hz in 1 Hz steps) and the signal from the microphone recorded. This was repeated at each arm.

To determine the mode dispersion, the temporal acoustic signals were summed at each arm producing a signal (voltage) as a function of propagation distance and time $V(x,t)$. The data were processed using temporal and spatial Fourier transform with a Tukey window function applied to determine the wavenumber-frequency relation.

\subsection{Acknowledgements}
The authors thank Edward Lupton from www.meccanospares.com for useful discussions on the Meccano\textsuperscript{\texttrademark} parts used throughout. G.J.C gratefully acknowledges financial support from the Royal Commission for the Exhibition of 1851 in the form of a Research Fellowship. T.A.S and A.P.H gratefully acknowledges financial support from DSTL. I.R.H acknowledges financial support from the EPSRC and QinetiQ Ltd. via the TEAM-A Prosperity Partnership (Grant No. EP/R004781/1), and from the EPSRC via the A-Meta project (Grant No. EP/W003341/1)


%

\clearpage
\onecolumngrid
\setcounter{figure}{0}
\renewcommand{\thefigure}{S\arabic{figure}}
\section{Supplemental Material}
\maketitle

\subsection*{Toy Model Details} 
\begin{figure}
\centering
\includegraphics[width=\textwidth]{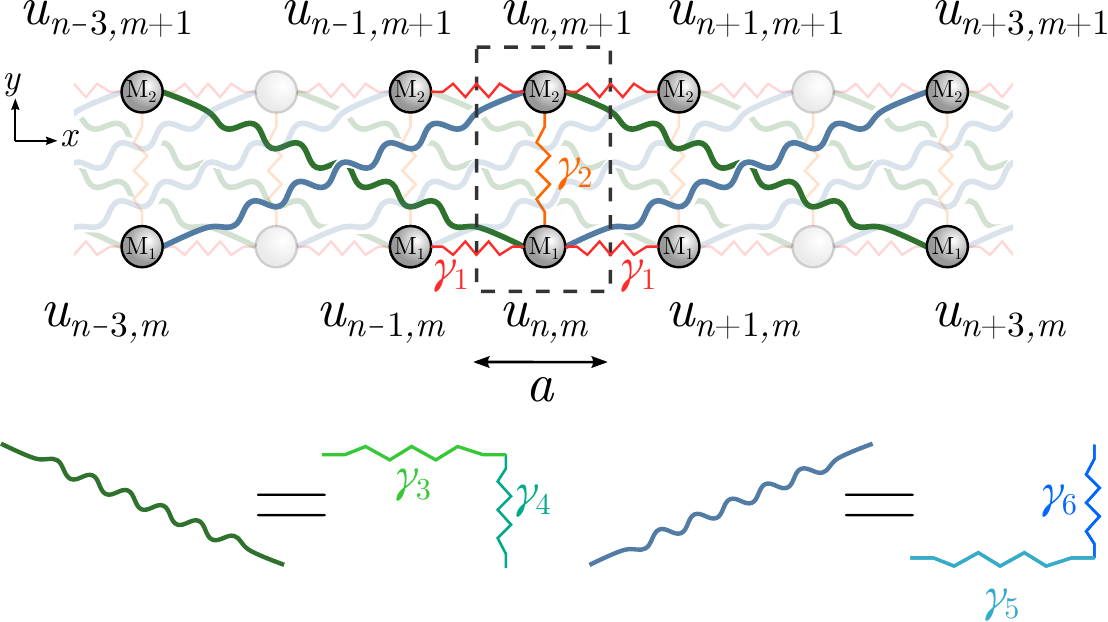}
\caption{Schematic of the toy model: Two coupled chains of masses ($M_1$ and $M_2$) with nearest and next-next-nearest coupling via the springs with spring constants $\gamma_i$, for $i = 1,\dots ,6$. The diagonal springs are modelled as two orthogonal springs that are restricted to move in $x$ and $y$ respectively. The unit cell of width $a$ is shown by the dashed rectangle. }
\label{fig:suppToy}
\end{figure}

We consider the forces exerted on the masses in the periodic unit cell marked by the dashed box in Fig.~\ref{fig:suppToy} arising from the classical Hookean response of the springs when the masses are displaced by $u_{n,m}$, shown in Fig.~\ref{fig:suppToy}. We model the diagonal connecting springs as two orthogonal springs that only displace in either the $x$ or $y$ directions. Solving for propagating Floquet-Bloch waves of frequency $\omega$, in the $x$ direction with wavenumber $k_x$, requires
\begin{equation}
    \begin{split}
    u_{n,m} &= U_{n,m}\exp{i(\omega t - k_xan)} \\ 
    u_{n+p,m} &= U_{n,m}\exp{i(\omega t - k_xa(n+p))},
    \end{split}
\end{equation}
where $a$ is the unit cell width, $p \in \mathbb{Z}$ and $U_{n,m}$ is the Bloch amplitude. Similar expressions for the masses in the second chain are described by displacements $u_{n,m+1}$. Considering the force $\boldsymbol{F} = (F_x,F_y)$ on the masses with displacements $u_{n,m}$ and $u_{n,m+1}$, and utilising Newton's third law, we arrive at a set of difference equations that can be solved by the eigenvalue problem $\left(\boldsymbol{A} - \omega^2\boldsymbol{M}\right)\boldsymbol{U} = 0$, i.e.
\begin{equation}
    \left(\underbrace{\begin{bmatrix}
    2\gamma_1(1-\cos(k_xa))+\Gamma & -\left(\Upsilon e^{3ik_xa} + \Lambda e^{-3ik_xa} + \gamma_2\right) \\
    -\left(\Lambda e^{3ik_xa} + \Upsilon e^{-3ik_xa} + \gamma_2\right) & 2\gamma_1(1-\cos(k_xa))+\Gamma
    \end{bmatrix}}_{\boldsymbol{A}}
    -\omega^2\underbrace{\begin{bmatrix}
    M_1 & 0 \\
    0 & M_2
    \end{bmatrix}}_{\boldsymbol{M}}\right)
    \underbrace{\begin{bmatrix} 
    U_{n,m} \\ U_{n,m+1}
    \end{bmatrix}}_{\boldsymbol{U}} = 0,
\end{equation}
where $\Gamma = \sum\limits_{i=2}^{6}\gamma_{i}$, $\Upsilon = \left(\gamma_3 + \gamma_4\right)$ and $\Lambda = \left(\gamma_5 + \gamma_6\right)$. The eigenvalue problem is solved numerically and the dispersion curves are obtained by iterating the wavenumber through the first Brillouin Zone, up to $k_x = \pi/a$. Tuning the parameters of the masses and springs replicates the effects of symmetry breaking of the Meccano\textsuperscript{\texttrademark} metamaterial presented in the main text. 

\begin{figure}
\centering
\includegraphics[width=\textwidth]{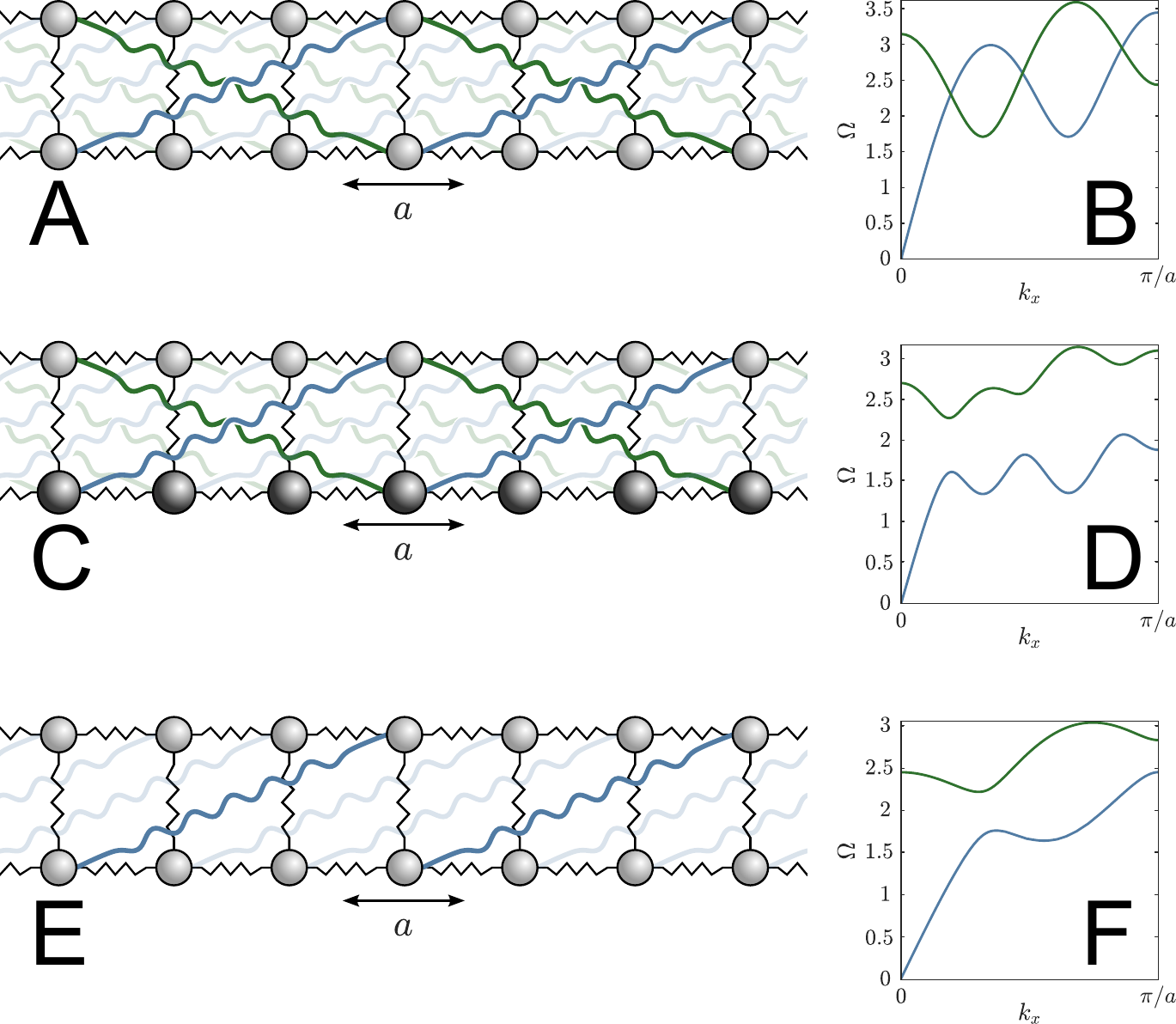}
\caption{Dispersion curves obtained from the toy models that qualitatively represent the results of the Meccano\textsuperscript{\texttrademark} structures in the main text: (A,B) schematic and dispersion curves of geometry representative of the symmetric configuration respectively. The parameters are such that $M_1 = M_2 = 1$ and all spring constants equal, $\gamma_i = 1$. (C,D) schematic and dispersion curves of geometry representative of the asymmetric configuration respectively. The two masses are now not equal, $M_1 = 2M_2$ ($M_2 = 1$), emulating the effect of displacing the central beam. (E,F) schematic and dispersion curves of geometry representative of the parallel configuration. The parameters are such that $M_1 = M_2 = 1$ and spring constants $\gamma_{1,2,5,6} = 1$ with $\gamma_3 = \gamma_4 = 0$.}
\label{fig:suppDisp}
\end{figure}

Figure~\ref{fig:suppDisp} shows three analogous set-ups of the toy model which replicate the dispersive properties of the Meccano\textsuperscript{\texttrademark} device. In Fig.~\ref{fig:suppDisp}(A,B) we show the schematic and dispersion curves respectively of the toy model presented in the main text, that emulates the dispersion of the symmetric configuration of the Meccano\textsuperscript{\texttrademark} structure. Two branches are visible with crossings due to the orthogonality of the normal modes, with similar `roton-like' character to that described in \cite{chen2021roton}, owing to the BNN coupling. In Fig.~\ref{fig:suppDisp}(C,D) we show a similar schematic and corresponding dispersion curves respectively, for a toy-model analogue of the asymmetric structure; the masses in the upper and lower chains are now not equal ($M_{1} = 2M_{2}$), thereby breaking mirror symmetry about $x$ in an analogous way to displacing the perpendicular arms in the asymmetric configuration of the Meccano\textsuperscript{\texttrademark} device, that lifts the degeneracies in the dispersion curves. In Fig.~\ref{fig:suppDisp}(E,F) we show the final toy model example that is representative of the parallel configuration; here only one diagonal connecting spring is included ($\gamma_3 = \gamma_4 = 0$). As in the analogous parallel Meccano\textsuperscript{\texttrademark} configuration, reflectional symmetry about $y$ here is broken, and similar distinct bands in the dispersion relation are observed. We note that the upper branches (green) in the toy models (Figs.~1, S2) do not approach zero frequency as $k_x \rightarrow 0$ as the model does not include phase-velocity limiting terms associated with the bulk compressional/transverse wavespeeds \cite{maznev2015waveguiding}.

\begin{figure}
\centering
\includegraphics[width=\textwidth]{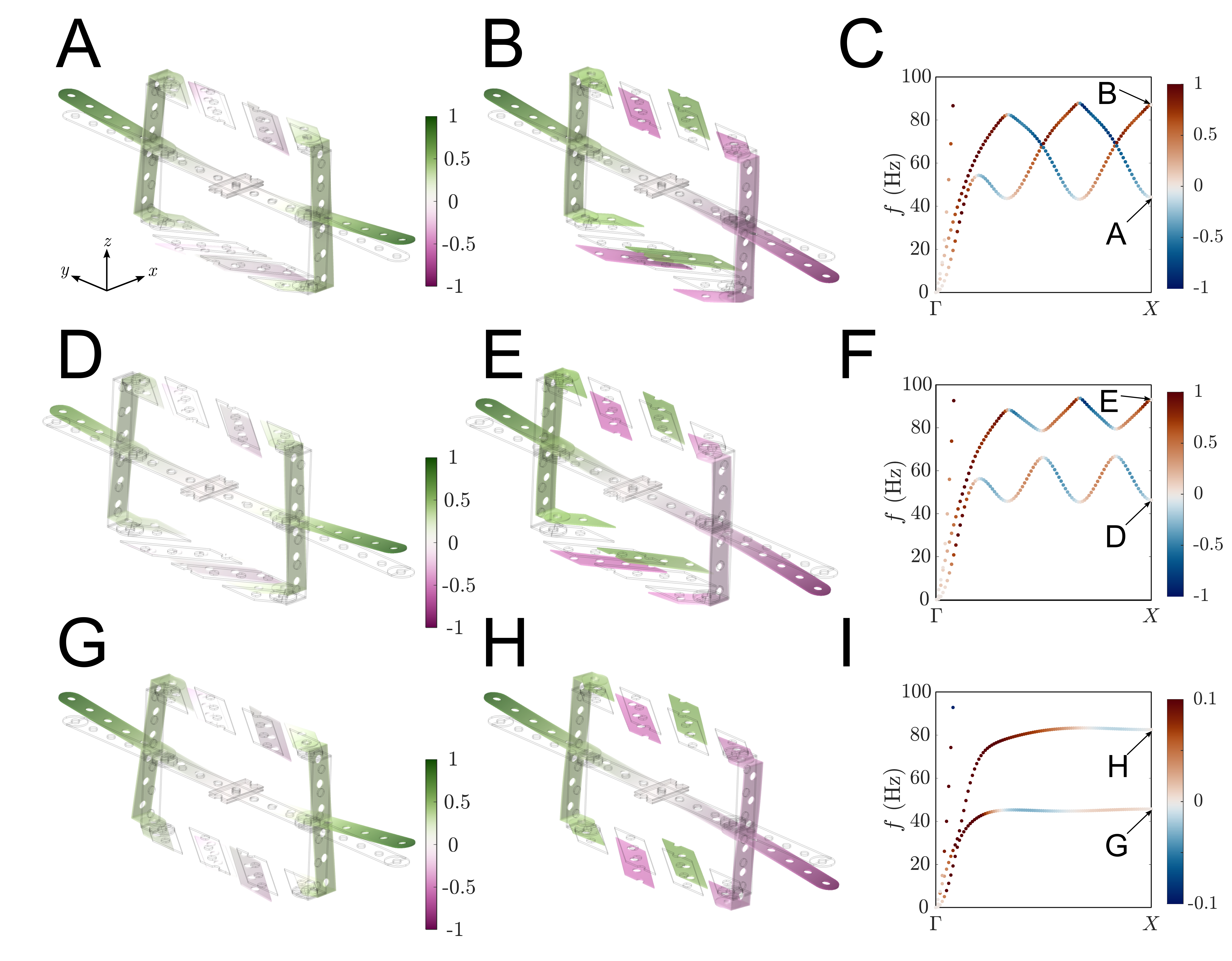}
\caption{Eigenmodes of FEM computations for the three configurations with colourscale showing the normalised out-of-plane displacement, $u_z$, with undistorted structure overlaid: (A,B) Eigenmodes of the symmetric configuration, showing the out-of-plane displacement of the perpendicular arm to be purely even or odd, respectively, evaluated at the BZ boundary - highlighted on the dispersion curves as in Fig.~2. (C) Eigenfrequency dispersion, as in Fig.~2(B), but with colourscale now showing the normalised net integrated complex mechanical energy flux over the whole unit cell, $\int_{cell}\tilde{I}_x dS$ (i.e. the sum of the integrals used in Fig.~2), showing that the net power flow is always in one direction, with that direction being determined by the gradient (positive or negative) of the dispersion relation. (D,E) as in (A,B) but for the asymmetric configuration, where $u_z$ of the perpendicular arm for each mode is neither purely odd or even. (F) as in (C) but for the asymmetric configuration. (G,H) as in (A,B) but for the parallel configuration. (I) as in (C) but for the parallel configuration.}
\label{fig:suppModes}
\end{figure}

\begin{figure}
\centering
\includegraphics[width=\textwidth]{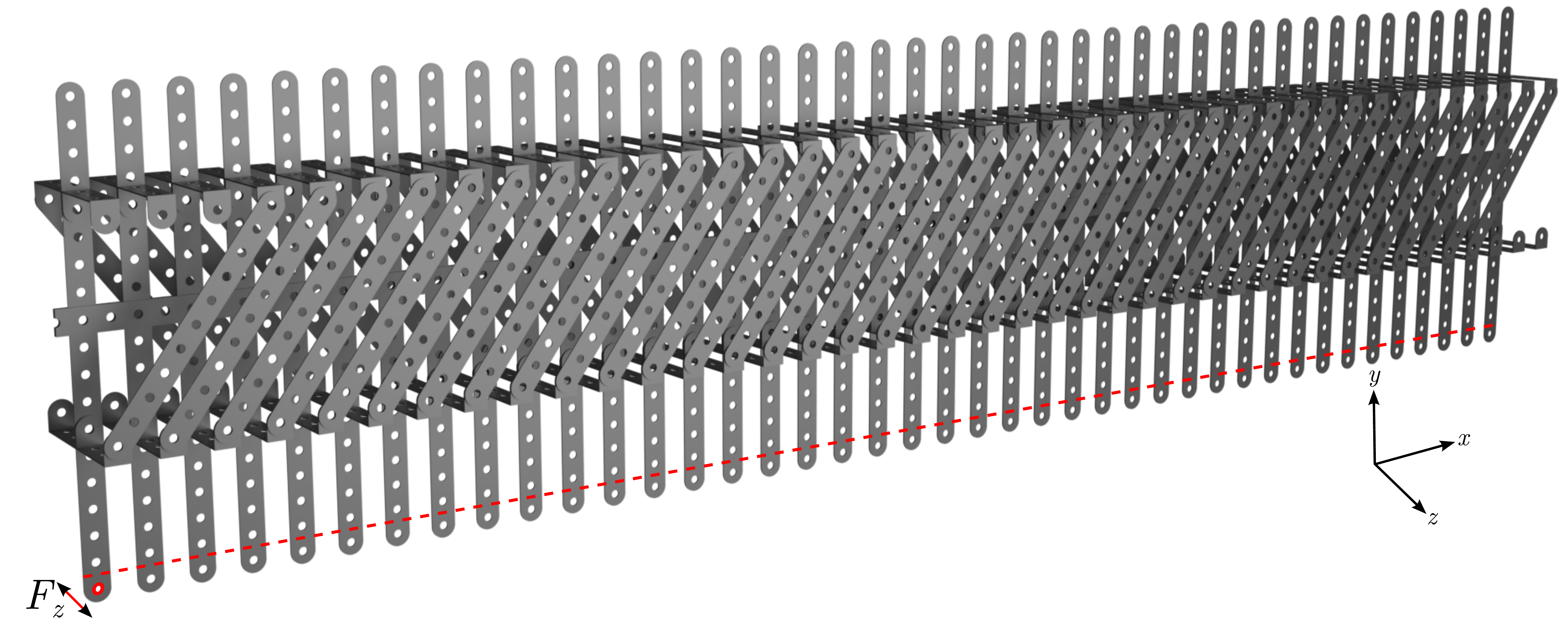}
\caption{FEM simulation domain (asymmetric configuration): Boundary load position shown by red circle where harmonic force $F_z$ is applied in the frequency domain simulations. The dashed red line shows the points where the out-of-plane displacement, $u_z$, was evaluated and subsequently spatially Fourier transformed to produce the FEM dispersion spectra on the left sides of Fig.~4.}
\label{fig:suppComsolSchem}
\end{figure}

\end{document}